\documentstyle[epsfig,aps,twocolumn]{revtex}
\begin{document}

\draft

\title{Quantum chaos in an ion trap: the delta-kicked harmonic oscillator}

\author{S.~A.~Gardiner, J.~I.~Cirac, and P.~Zoller }
\address{Institut f{\"u}r Theoretische Physik,
Universit{\"a}t Innsbruck, 6020
Innsbruck, Austria}
\date{\today}
\maketitle

\begin{abstract}
We propose an experimental
configuration, within an ion trap, by which a quantum mechanical delta-kicked
harmonic oscillator could be realized, and investigated. We show how to 
directly measure the sensitivity of the ion motion to small variations 
in the external parameters.
\end{abstract}



In classical mechanics, deterministic chaos is often most simply described as
{\em exponential sensitivity to initial conditions}, meaning that initially
neighbouring classical trajectories diverge extremely rapidly with time.
Due to the necessity of preserving the inner product, this kind of divergence 
between two possible initial states cannot occur quantum mechanically. 
The question of what then constitutes the quantum mechanical equivalent of chaos
immediately arises. An interesting proposal by Peres \cite{peres} is to examine
the initial state $|\psi\rangle$ evolving
under two slightly differing (classically chaotic) Hamiltonians $\hat{H}_{1}$
and $\hat{H}_{2}$. Defining $U_{1,2}(t)$ as the corresponding unitary
evolution operators,
the overlap
\begin{equation}
O =
\left|\langle\psi|
U_2(t)^\dagger U_1(t)
|\psi\rangle
\right|^{2}
\label{overlap}
\end{equation}
is predicted to behave very differently, depending on whether the initial state
is in a stable or chaotic area of phase space. Thus $O$ is a measure to
distinguish between regular and irregular quantum dynamics. In fact much
work on the subject of quantum chaos has been carried out theoretically; 
experimental realizations however remain somewhat 
scarce \cite{misc}, although
there have  recently been pioneering successes in atom optics \cite{raizen} and 
in mesoscopic solid state systems \cite{fromhold}.
In this Letter we propose a realizable experimental 
configuration, with which one can measure $O$ directly. The system
proposed is a single ion trapped in a harmonic potential, subject to periodic 
kicks from a standing wave laser. This is a quantum delta-kicked harmonic
oscillator, a system capable classically 
of stochastic dynamics, including Arnol'd diffusion
\cite{arnold} under certain resonance conditions \cite{chernikov}.

Trapped ions are in many ways an ideal choice of system for the study of
fundamental aspects of quantum mechanics. One can take advantage of
the small dissipation in this system, together with the
possibility of {\em coherent\/} manipulation of the ion's motional
state. Trapped ions have been used in recent experimental demonstrations
of the generation of non-classical states of motion \cite{winelandcats},
quantum logic gates \cite{winelandgates}, and tomography of the density
matrix \cite{winelandtomography}. In addition there have been
theoretical proposals for investigation of localization
\cite{schleich}
and irregular collapse and revival dynamics \cite{milburn}
due to quantum chaos 
in this system.

In this Letter we will first describe a general procedure for determining 
$O$ as a function of time. 
We then describe explicitly a particular system, the
delta-kicked harmonic oscillator, how it may be implemented within an ion trap,
and how to carry out our general procedure for determining $O$.
 Finally we display some numerical results, showing what one
would expect to see when carrying out such an experiment. Markedly different
results are indeed observed numerically, dependent on whether the initial
condition is in a classically stable or chaotic area of phase space.

We first consider a general Hamiltonian of the form
$
\hat{H} = \hat{H}_{1}|g_{1}\rangle\langle g_{1}| + 
\hat{H}_{2}|g_{2}\rangle\langle g_{2}|,
$
where $|g_{1}\rangle$ and ¼$|g_{2}\rangle$ are stable
electronic ground states of a single trapped
ion. The state of the ion is set initially to be:
\begin{equation}
|\psi(0)\rangle = 
\frac{1}{\sqrt{2}}\left(|g_{1}\rangle|\alpha \rangle + |g_{2}\rangle|\beta
\rangle\right). 
\label{initial}
\end{equation}
Here $|\alpha\rangle$ and $|\beta\rangle$ are states of the ion's motion.
In the applications described in this Letter, 
$|\alpha\rangle$ and $|\beta\rangle$ are coherent states; 
§$|\psi(0)\rangle$ is then a Schr\"{o}dinger cat state,
which has been achieved experimentally for a trapped ion \cite{winelandcats}.
After a time $t$, the initial state evolves to
$
|\psi(t)\rangle = 
[|g_{1}\rangle\hat{U}_{1}(t)|\alpha \rangle + 
|g_{2}\rangle\hat{U}_{2}(t)|\beta
\rangle]/\sqrt{2},
$
where $\hat{U}_{1}(t)$ and $\hat{U}_{2}(t)$ are the time evolution operators 
derived from $\hat{H}_{1}$ and $\hat{H}_{2}$, respectively. A $\pi/2$ pulse is
applied (Ramsey type experiment) to the ion, yielding
\begin{eqnarray}
|\psi(t)\rangle' &=& \frac{1}{2}\left\{
|g_{1}\rangle\left[\hat{U}_{1}(t)|\alpha \rangle -
\hat{U}_{2}(t)|\beta \rangle\right] + \right. \label{pipulse}\\
&& \left.
|g_{2}\rangle\left[\hat{U}_{1}(t)|\alpha \rangle +
\hat{U}_{2}(t)|\beta \rangle\right]\right\}.
\nonumber
\end{eqnarray}
The probability for the ion to be in state $|g_{1}\rangle$ is thus
\begin{equation}
P_{g} = \frac{1}{2}\left\{1 - \mbox{Re}\left[\langle
\beta|\hat{U}_{2}^{\dagger}(t)\hat{U}_{1}(t)|\alpha\rangle
\right]\right\}.
\label{Pg}
\end{equation} 
Similarly, if we set 
$
|\psi(0)\rangle =
(|g_{1}\rangle|\alpha\rangle + i|g_{2}\rangle|\beta\rangle)/\sqrt{2},
$
the corresponding final probability is given by
\begin{equation}
P_{g}' = \frac{1}{2}\left\{1 - \mbox{Im}\left[\langle
\beta|\hat{U}_{2}^{\dagger}(t)\hat{U}_{1}(t)|\alpha\rangle
\right]\right\}.
\label{Pgprime}
\end{equation}

By determining $P_{g}$ and $P_{g}'$, one can clearly deduce
$
|\langle \beta |\hat{U}_{2}^{\dagger}(t)\hat{U}_{1}(t)|\alpha 
\rangle|^{2}.
$
If $\hat{H}_{1}$ and $\hat{H}_{2}$ are slightly differing chaotic Hamiltonians,
and $|\alpha\rangle = |\beta\rangle$, then we have $O$, as defined
in Eq.\ (\ref{overlap}). We also note that
where $\hat{H}_{2}$ is the simple harmonic oscillator Hamiltonian, this
collapses to $2\pi Q(\beta)$, where $Q(\beta)$ is the $Q$ function for various
initial $\beta$ of the pure state $\hat{U}_{1}(t)|\alpha\rangle$. By repeated
measurements one can therefore determine the $Q$ function's evolution in time
\cite{ionwigner}.

\begin{figure}[htbp]
\begin{center}\
\epsfig{file=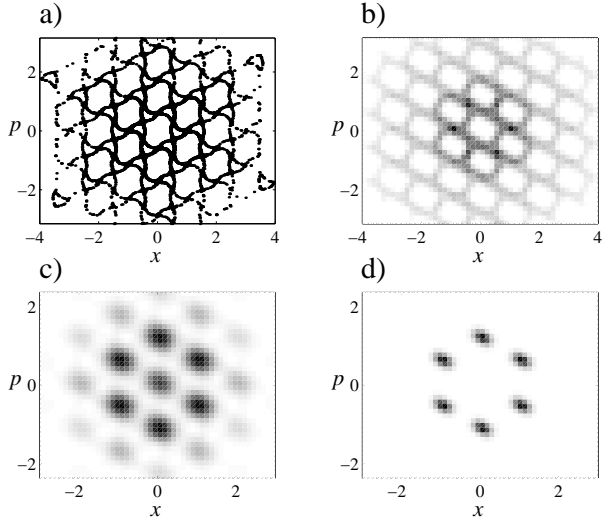,width=80mm}
\end{center}
\caption{
Stroboscopic Phase space portraits (from kick to kick)
for the delta-kicked harmonic oscillator, where $r/q =
1/6$ and $\kappa = 0.2$, after averaging over $11000$ kicks.
The position $x$ and momentum $p$
axes are in units of $\lambda$ and $m\nu\lambda$ respectively.
(a) Poincar\'{e} surface of section, showing the classical stochastic web. 
Regular dynamics take place within the cells defined by the web. 
(b) Time averaged $Q$ function for the (unstable) initial condition 
$|\alpha\rangle$ where
$\alpha = \pi/2\eta$ and $\eta = 0.25$, centred at $(1,0)$.
(c) Time averaged $Q$ function for the (stable) initial condition 
$\alpha = i\pi/\eta{\protect \sqrt{3}}$ ($(0, 2/{\protect \sqrt{3}})$) and 
$\eta =0.5$.
(d) As for (c), where $\eta = 0.25$. Note the ``tunneling'' out of
the original ring of cells
for the larger value of $\eta$.
}
\label{webs}
\end{figure}

Our proposed model system is a
harmonic oscillator
\begin{equation}
H_{0} = \frac{p^2}{2m} + \frac{m\nu^2 x^2}{2},
\label{harmham}
\end{equation} 
periodically perturbed by nonlinearly position-dependent delta-kicks;
\begin{equation}
H_1 =
K\cos(kx)\sum_{n = -\infty}^{\infty}\delta(t-n\tau),
\label{classham}
\end{equation}
so that $H = H_{0} + H_1$.
Here $x$ is the position,
$p$ the momentum,
 $m$ the mass, $\nu$ the oscillator frequency,
$k = 2\pi/\lambda$ the wavenumber, $t =$ time,
$\tau
$ the time delay between the kicks, and
$K$ the kick strength.
Under the resonance condition $ß\nu\tau = 2\pi r/q$ ($r/q$ is a positive
rational, where $q>2$),
classically there are thin channels of chaotic dynamics in the
phase space \cite{chernikov}. The resulting 
Arnol'd  stochastic web [see Fig.\ \ref{webs}(a)] spreads through all of phase 
space; Arnol'd diffusion \cite{arnold} can occur in systems of less than two
dimensions when the conditions for the KAM (Kolmogorov, Arnol'd, Moser) 
theorem \cite{kam} are not fulfilled, as is the
case here \cite{chernikov}. 
The corresponding quantum mechanical system has also
been studied theoretically \cite{quantkick} [see Figs.
\ref{webs}(b,c,d) for the 
time averaged $Q$ function of this system].

\begin{figure}[htbp]
\begin{center}\
\epsfig{file=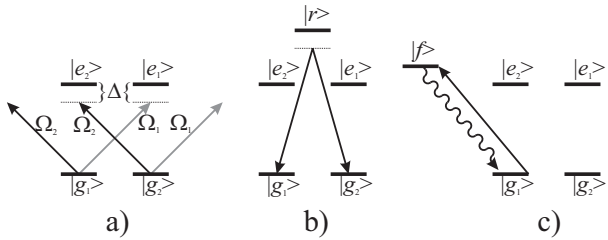,width=80mm}
\end{center}
\caption{Atomic level configuration proposed for our procedure. 
(a) The atom first experiences a series of short laser pulses from an 
elliptically polarized standing wave. Levels $|g_{1}\rangle$ and $|e_{1}\rangle$ 
are coupled by $\sigma_{+}$polarized light, and levels $|g_{2}\rangle$ and 
$|e\rangle$ by $\sigma_{-}$ polarized light, where the corresponding intensities 
differ. 
(b) After a definite number of kicks, 
the levels $|g_{1}\rangle$ and $|g_{2}\rangle$ experience a $\pi/2$ pulse 
({\em e.g}.\ a Raman transition, using the auxiliary level $|r\rangle$).
(c) The population of $|g_{1}\rangle $ is
determined by fluorescence, using the auxiliary level $|f\rangle$.
}
\label{atoms}
\end{figure}

To construct such a system quantum mechanically, 
which can also be used to carry out the procedure described in 
Eqs.~(\ref{pipulse},\ref{Pg},\ref{Pgprime}), 
we begin with a single 
ion in a harmonic potential ({\em e.g}.\ a linear ion trap \cite{winelandcats});
in addition we require
a time dependent
standing wave laser configuration. The ion has two ground states and two excited
states, and the laser is elliptically polarized [see
Fig.\ \ref{atoms}(a)].
The $\sigma_{+}$ and $\sigma_{-}$ polarized contributions 
thus separately couple
two different two level systems, with different Rabi frequencies:
\begin{eqnarray}
\hat{H} &=& 
\hat{H}_{0}+
\frac{\hbar}{2}\sum_{j=1}^{2}\left\{
\omega_{0}(|e_{j}\rangle\langle e_{j}| - |g_{j}\rangle\langle 
g_{j}|)
\right.\\&&\left.
+\cos(k\hat{x})
\left[\Omega_{j}(t)
e^{-i\omega_{L}t}|e_{j}\rangle\langle g_{j}| + \mbox{H.c.}\right]\right\}, 
\nonumber
\end{eqnarray}
where $\omega_{0}$ is the transition frequency between the electronic states
$|e_{j}\rangle$ and $|g_{j}\rangle$, $\omega_{L}$ is the laser frequency, and 
$\Omega_{1,2}(t)$
 are the (time dependent) Rabi frequencies. In a rotating frame defined
by $\hat{U} =
\exp[-i\omega_{L}t\sum_{j=1}^{2}
(|e_{j}\rangle\langle e_{j}|-
|g_{j}\rangle\langle g_{j}|)/2]$, 
and in the limit
of large detuning $|\Delta| = |\omega_{L} - \omega_{0}|\gg |\Omega_{1,2}(t)|$, 
$|e_{1}\rangle$ and $|e_{2}\rangle$ can be
adiabatically eliminated to give:
\begin{equation}
\hat{H} = 
\hat{H}_{0}+
\frac{\hbar}{8\Delta}\sum_{j = 1}^{2}\Omega_{j}(t)^{2}
\left[ \cos \left(2k\hat{x} \right)+1\right]
|g_{j}\rangle\langle g_{j}|.
\end{equation}
The laser is rapidly and periodically switched, giving a series of short
Gaussian pulses:
\begin{equation}
\Omega_{j}(t)^{2} = \Omega_{j}^{2}\sum_{n=-\infty}^{\infty} e^{-(t -
n\tau)^{2}/\sigma^{2}}, 
\end{equation}
which approximate a series of delta kicks in the limit $\sigma\rightarrow 0$.
Note also that we require $\sigma \gg 1/\Delta$, otherwise the laser is too
spectrally broad, making adiabatic elimination of $|e_{1}\rangle$ and
$|e_{2}\rangle$ impossible.
Thus, finally, we have:
\begin{equation}
\hat{H} = 
\hat{H}_{0}+
\sum_{j=1}^{2}K_{j}
\left[\cos\left(2k\hat{x}\right)+1\right]
|g_{j}\rangle\langle g_{j}|
\sum_{n =-\infty}^{\infty}\delta(t-n\tau),
\end{equation}
which corresponds almost exactly to Eqs.~(\ref{harmham},\ref{classham}), 
for two different $K_{j} = \hbar\sigma\sqrt{\pi}\Omega_{j}^{2}/8\Delta$. There 
are extra $|g_{j}\rangle\langle g_{j}|$ terms, but these will only contribute 
phases to the evolution of the initial state of Eq.\ (\ref{initial}), and can 
easily be accounted for. 

Taking $|\alpha\rangle=|\beta\rangle$, the initial state of Eq.\ (\ref{initial}) 
thus evolves as
\begin{equation}
|\psi(n\tau)\rangle = \frac{1}{\sqrt{2}}
\sum_{j=1}^{2}
e^{-in\kappa_{j}/\sqrt{2}{\eta}^{2}}\hat{F}_{j}^{n}|g_{j}\rangle|\alpha\rangle
\end{equation}
where the Floquet time evolution operators $\hat{F}_{j}$ are given by:
\begin{equation}
\hat{F}_{j} = 
e^{-i\hat{a}^{\dagger}\hat{a}\nu\tau}e^{-i\kappa_{j}\cos[2\eta(\hat{a}^{\dagger}+
\hat{a})]|g_{j}\rangle\langle g_{j}|/\sqrt{2}{\eta}^{2}},
\label{realflo}
\end{equation}
where $\hat{a}^{\dagger}$ and $\hat{a}$ respectively create and
annihilate a single phonon quantum, and the common phase term $e^{-i\nu\tau/2}$ has
been dropped.
The $\kappa_{j} = \Omega_{j}^{2}{\eta}^{2}\sigma\sqrt{2\pi}/8\Delta$ 
are dimensionless kick parameters, which, with $\nu\tau$, determine fully the
phase space behaviour of the classical delta-kicked harmonic oscillator
\cite{chernikov}. 
In the quantum mechanical problem there is 
an additional
parameter, the
 Lamb-Dicke parameter $\eta =
k\sqrt{\hbar/2m\nu}$.
 As $\eta^2 \propto \hbar$, by 
progressively reducing $\eta$, one can explore the transition from quantum to 
classical chaos \cite{loc}. This can be accomplished by ``tightening'' or ``loosening''
the trapping potential, {\it i.e}.\ increasing or decreasing the trapping frequency
$\nu$. 
 
After $n$ kicks, we perform a $\pi/2$ pulse between the levels
$|g_{1}\rangle$ and $|g_{2}\rangle$, by {\em e.g}.\ a Raman transition or a
magnetic field [see Fig.\ \ref{atoms}(b)].
By fluorescence, using an auxiliary level $|f\rangle$, with repeated
measurements one can 
determine
 $P_{g}$ and $P_{g}'$ [see Fig.\ \ref{atoms}(c)], as
defined in Eqs.~(\ref{Pg},\ref{Pgprime}).
\begin{eqnarray}
P_{g} & = & \frac{1}{2}\left[
1 -
\cos(\delta\kappa n/\sqrt{2}{\eta}^{2})
\mbox{Re}\left(\langle\alpha|\hat{F}_{2}^{\dagger n}\hat{F}_{1}^{n}|\alpha\rangle\right) -
\right.\label{preolap}\\ &&\left.
\sin(\delta\kappa n/\sqrt{2}{\eta}^{2})
\mbox{Im}\left(\langle\alpha|\hat{F}_{2}^{\dagger n}\hat{F}_{1}^{n}|\alpha\rangle\right)
\right]
\nonumber\\
P_{g}' & = & \frac{1}{2}\left[
1 -
\sin(\delta\kappa n/\sqrt{2}{\eta}^{2})
\mbox{Re}\left(\langle\alpha|\hat{F}_{2}^{\dagger n}\hat{F}_{1}^{n}|\alpha\rangle\right) -
\right.\nonumber\\&&\left.
\cos(\delta\kappa n/\sqrt{2}{\eta}^{2})
\mbox{Im}\left(\langle\alpha|\hat{F}_{2}^{\dagger n}\hat{F}_{1}^{n}|\alpha\rangle\right)
\right],
\nonumber
\end{eqnarray}
where $\delta\kappa = \kappa_{2} - \kappa_{1}$. From Eq.\ (\ref{preolap})
one can easily extract the overlap 
$O=|\langle\alpha|\hat{F}_{2}^{\dagger n}\hat{F}_{1}^{n}|\alpha\rangle|^{2}$.

In order to relate the quantum mechanical behavior of the system to the
classical one, we have to use an initial condition equivalent to the 
classical $x(0)$ and $p(0)$. We use a coherent state
$|\alpha\rangle$, where $\alpha$ can be expressed as
$\alpha = [kx(0) +ikp(0)/m\nu]/4\eta$.
Thus we can see that when $\eta$ is small, $\alpha$ is large, and
$|\alpha\rangle$ is therefore more macroscopic, in some sense more {\em
classical}. This can be seen by comparing Fig.\ \ref{webs}(c) with Fig.\
\ref{webs}(d); for $\eta = 0.5$ population ``tunnels'' through a classically
forbidden area, which does not occur when $\eta = 0.25$ \cite{tunnelnote}.

\begin{figure}[htbp]
\begin{center}\
\epsfig{file=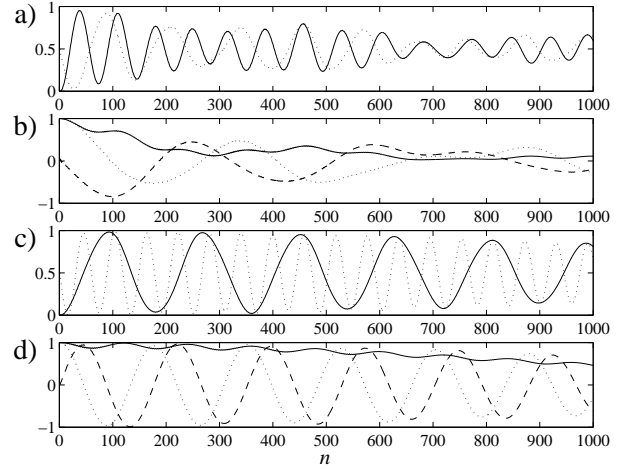,width=80mm}
\end{center}
\caption{
(a) $P_{g}$ (solid line) and $P_{g}'$ (dotted line) for the initial condition
$\alpha = \pi/2\eta$  ($\eta = 0.5$, $\kappa_{1} = 0.2$, $\kappa_{2} = 0.225$).
(b) $O$ (solid line), $\mbox{Re}(\langle \alpha| \hat{F}_{-}^{n\dagger}
\hat{F}_{+}^{n}|\alpha\rangle)$ (dashed line), and
$\mbox{Im}(\langle \alpha| \hat{F}_{-}^{n\dagger}
\hat{F}_{+}^{n}|\alpha\rangle)$ (dotted line) for the same initial
condition.
(c) and d), same as a) and b), for the initial condition $\alpha =
i\pi/\eta{\protect \sqrt{3}}$.  
}
\label{olone}
\end{figure}

Figure \ref{olone} shows the values of $P_{g}$ and $P_{g}'$ that one would 
measure for this scheme, and 
the value of $O$ that one would thus obtain, for $\eta = 0.5$ after $0$--$1000$
kicks.
The plots obtained are clearly different, depending on whether the initial
condition is classically unstable, as in Figs.\ \ref{olone}(a,b) where 
the corresponding classical initial condition is a hyperbolic fixed point, or 
stable
[Figs.\ \ref{olone}(c,d), elliptic fixed point]. This is already
noticeable in the plots of $P_{g}$ and $P_{g}'$, before $O$ is extracted 
[Figs.\ \ref{olone}(a,c)]. 
In line
with previous numerical work for the kicked top \cite{peres}, $O$ decays for an
unstable initial condition, and undergoes quasistable oscillations for a stable
initial condition. 

As $O$ is a measure of how close the two parallel
evolutions are at a given time, it can be seen that if the initial condition is
classically unstable [Fig.\ \ref{olone}(b)] the two states become 
rapidly increasingly orthogonal (more ``far apart''), 
whereas in the case of a stable initial condition, for some time the
difference between the states remains on average about the same. This in some
sense corresponds to the classical definition of chaos, where under the
influence of the {\em same\/} dynamics, very slightly {\em different\/} initial
states diverge rapidly if their origin is in an unstable area of phase space
\cite{peres}.

\begin{figure}[htbp]
\begin{center}\
\epsfig{file=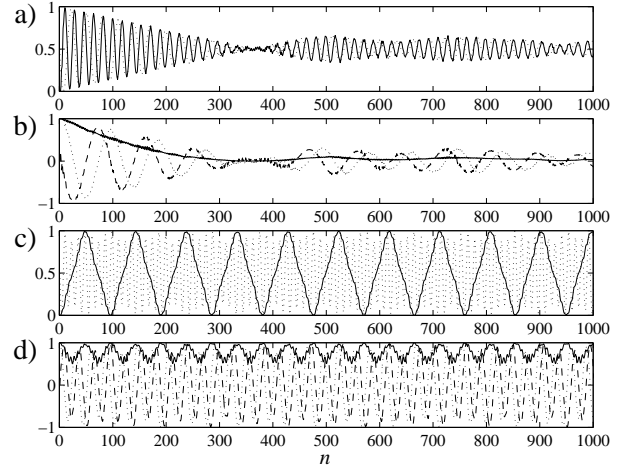,width=80mm}
\end{center}
\caption{
Corresponds exactly to Fig.\ \protect{\ref{olone}}, except that $\eta = 0.25$
}
\label{olhalf}
\end{figure}

Figure \ref{olhalf} shows the same for $\eta = 0.25$. In line with the fact that
this is more in the semiclassical regime than Fig.\ \ref{olone}, the decay 
[Figs.\
\ref{olhalf}(a,b)] is more rapid, and the oscillations 
[Figs.\ \ref{olhalf}(c,d)]
are more stable. The slow decay of the quasistable oscillations
when $\eta=0.5$ [Fig.\ \ref{olone}(d)] can be traced back to the
tunneling that takes place in this regime (see Fig.\ \ref{webs}), absent when 
$\eta = 0.25$.

Numerically the
procedure is carried out in a truncated Fock basis of 400 states when 
$\eta=0.5$, or
800 when $\eta=0.25$. Increasing the size of the Fock basis does 
not qualitatively change the observed dynamics.

In conclusion we have shown a general procedure for determining the overlap
parameter $O$ originally proposed by Peres \cite{peres}.
We have described explicitly how $O$ could be determined for the
delta-kicked harmonic oscillator, a classically chaotic system. 
We have described how a single ion trapped in a harmonic
potential could be a practical experimental realization of the 
delta-kicked harmonic
oscillator, and how 
our scheme for determining $O$ is realized in this
configuration. In particular, our scheme presents a direct way for determining $O$, by virtue of
the fact that we effectively have two Hamiltonians running in parallel, within
the same experimental system.

We thank
R. Blatt,
J. Eschner,
P. Gerwinski,
F. Haake, 
J. P. Paz,
W. P. Schleich,
H. Schomerus,
P. T\"{o}rm\"{a},
D. J. Wineland, and
W. H. Zurek
for discussions.
This work was supported by 
the Austrian Fond zur F\"orderung der wissenschaftlichen Forschung
and
TMR network ERBFMRX-CT96-0002.

\begin{references}
\bibitem{tunnelnote}
The tunneling effect can be measured by tomographic methods 
(see \protect{\cite{winelandtomography,ionwigner}}).
One can also also observe the tunneling to the center of the web using
the present scheme with 
$\beta=0$ in Eq.~(\ref{initial}) and $\hat U_1 (t)$ the free evolution
of the harmonic oscillator [compare Eqs.~(\ref{Pg}) and (\ref{Pgprime})].
\end{references}
\end{document}